\documentclass[pre,twocolumn]{revtex4}
\usepackage{epsfig}
\usepackage{bm}
\usepackage{amsmath}
\usepackage{hyperref}
\usepackage{breakurl}

\begin{document}

\title{Chaos and turbulence in bubbly flows}

\author{A. Bershadskii}

\affiliation{
ICAR, P.O. Box 31155, Jerusalem 91000, Israel
}

\begin{abstract}

  Results of direct numerical simulations and laboratory experiments have been used in order to show that the buoyancy-driven bubbly flows at high gas volume fraction are mixed by deterministic chaos with a typical exponential spectrum of the liquid kinetic energy, whereas at moderate and small gas volume fraction it is a distributed chaos (turbulence or pseudo-turbulence) dominated by the third and second moments of helicity distribution with the stretched exponential spectra of the kinetic energy. Interaction of the bubbles with isotropic (behind an active grid) and near-wall turbulent flows has been also discussed from this point of view with an application to the pressurized water nuclear reactors.

\end{abstract}

\maketitle

\section{Introduction}

     The bubbly flows are usually rather complex - multiphase, buoyancy-driven, with different types of background turbulence, etc. Therefore, different classifications of these flows were used (see for recent reviews Refs. \cite{ris}-\cite{ppf} and references therein). \\

  The bubbly flows can be roughly divided, for instance, into two classes: `pseudo-turbulence' and `buoyancy-driven turbulence' \cite{ris}. In the `pseudo-turbulence' the collective (homogeneous) effects due to the rising of multiple single bubbles and their wakes at comparatively small gas fraction play the main role, whereas in the `buoyancy-driven turbulence' the effects of inhomogeneous swarms of rising bubbles (with the buoyancy-caused instabilities) at comparatively high gas fraction are the main factor. \\
  
   The turbulence can be transformed into deterministic chaos at the high gas fractions. Figure 1 shows (in the log-log scales) horizontal energy spectrum obtained in a recent direct numerical simulation at high gas volume fraction $\langle \alpha_g \rangle = 0.5$. The spectral data for the Fig. 1 were taken from figure 12b of the Ref. \cite{ppf}. The dashed curve is drawn to indicate the exponential spectrum
 $$
 E(k) \propto \exp-(k/k_c)   \eqno{(1)}
 $$
 where $k_c$ is a characteristic wavenumber (its position is indicated by the dotted arrow). We will return to this DNS with more details below.\\

\begin{figure} \vspace{-0.7cm}\centering
\epsfig{width=.45\textwidth,file=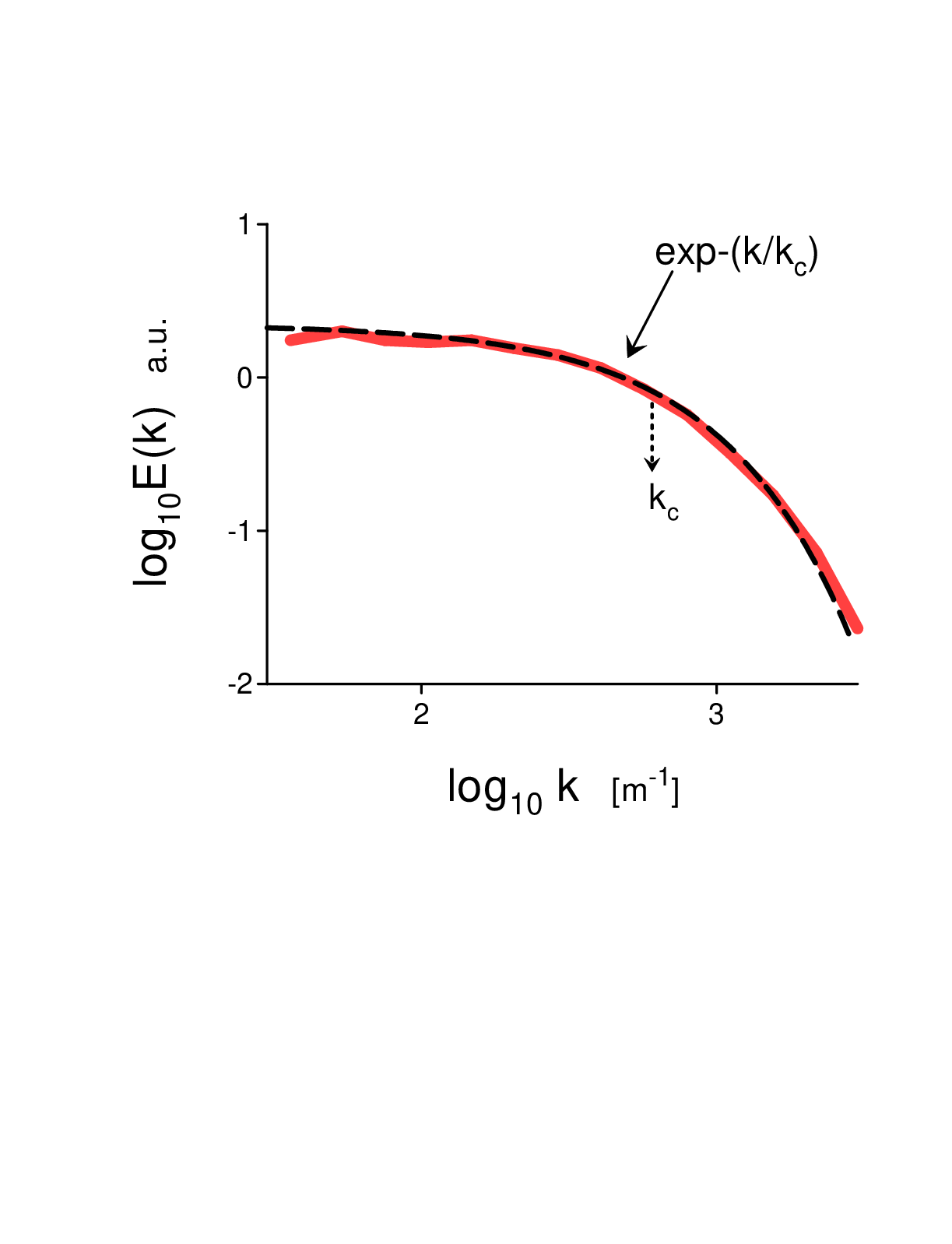} \vspace{-3.6cm}
\caption{Horizontal kinetic energy spectrum at high gas volume fraction $\langle \alpha_g \rangle = 0.5$.} 
\end{figure}
   
 The exponentially decaying spectra (in the wavenumber and frequency domains) are typical for the deterministic chaos \cite{oh}-\cite{kds}. The turbulence can be restored (through the distributed chaos) with decreasing in the gas volume fraction. Since the velocity field of the `buoyancy-driven turbulence' is characterized by significant anisotropy (see, for instance, the Ref. \cite{ppf} and references therein), the vertical energy spectrum can be dominated by the distributed chaos whereas the horizontal spectrum is already dominated by the deterministic chaos (cf. Fig. 4) due to the high value of the gas fraction.\\
     
     In Section II a possibility of the appearance of the helical distributed chaos, dominated by the third moment of the helicity distribution, in the bubbly flows has been presented and the kinetic energy spectrum typical for this chaos has been obtained. In Section III this spectrum has been compared with the spectra obtained in different numerical simulations and in a laboratory experiment. In Section IV  a general case of the helical distributed chaos has been considered and the kinetic energy spectra typical for this chaos have been obtained. In Section V these spectra have been compared with the spectra obtained in different numerical simulations and in a laboratory experiment.

\section{Helical distributed chaos}

   At the simplest approach to the problem of the bubbly flows the dynamics of the liquid phase is simulated using the Navier-Stokes equations for  incompressible viscous flow 
 $$ 
 \frac{\partial {\bf u}}{\partial t} + ({\bf u} \cdot \nabla) {\bf u}  =  -\frac{\nabla p}{\rho} + \nu \nabla^2 {\bf u}  +{\bf F}  \eqno{(2)}
$$
$$
\nabla \cdot {\bf u} =  0 \eqno{(3)}
$$
with the body force term ${\bf F}$ in the Eq. (2) exerted at the bubble-liquid interface (see, for instance, Ref. \cite{lai} and references therein).\\

     For the inviscid case ($\nu=0$) the equation for the mean helicity corresponding to the Eqs. (2-3) is
$$
\frac{d\langle h \rangle}{dt}  = 2\langle {\boldsymbol \omega}\cdot {\bf F}  \rangle \eqno{(4)} 
$$ 
where the helicity density is $h={\bf u}\cdot {\boldsymbol \omega}$, the vorticity is ${\boldsymbol \omega} = \nabla \times {\bf u}$ and $\langle...\rangle$ represents an average over the liquid volume. It is clear that the mean helicity cannot be generally considered as an inviscid invariant for these flows.  Let us consider the case when the large-scale motions contribute the main part to the correlation $\langle {\boldsymbol \omega}\cdot {\bf F}  \rangle$, but the correlation is quickly approaching zero with reducing spatial scales (that is typical for turbulent flows). Despite the mean helicity is not an inviscid invariant the higher moments of the helicity distribution $h={\bf u}\cdot {\boldsymbol \omega}$ can be considered as the inviscid invariants in this case \cite{mt},\cite{lt}.\\

   Indeed, let us divide the liquid volume into cells with volumes $V_i$ moving with the liquid (using the Lagrangian description) \cite{mt}\cite{lt}. An additional restriction on the considered cells is the boundary condition ${\boldsymbol \omega} \cdot {\bf n}=0$ on their surfaces $S_i$. Moment of order $n$ can be defined as 
 $$
I_n = \lim_{V \rightarrow  \infty} \frac{1}{V} \sum_j H_{j}^n  \eqno{(5)}
$$
with the helicity in the subvolume $V_i$
$$
H_j = \int_{V_j} h({\bf r},t) ~ d{\bf r}.  \eqno{(6)}
$$
   Due to the quick reduction of the correlation $\langle {\boldsymbol \omega}\cdot {\bf F} \rangle$ with the spatial scales the partial helicities $H_j$ can be still (approximately) considered as inviscid invariants for the cells $V_i$ where the spatial scales of the liquid motion are small enough. These cells provide the main contribution to the $I_n$ with $n \gg 1 $ for sufficiently chaotic (turbulent) flows (cf. \cite{bt}).  Hence, $I_n$ for sufficiently large $n$ can be considered as an inviscid quasi-invariant whereas the total helicity $I_1$ cannot. For sufficiently chaotic (turbulent) flows the value $n=3$ and even value $n=2$ can be regarded as sufficiently large  ($I_2$ is the Levich-Tsinober invariant of the Euler equation \cite{lt}). For the viscid cases, the `high' moments $I_n$ can be still regarded as adiabatic invariants in the inertial range of scales. \\ 

    Let us now return to Fig. 1. When the gas volume fraction becomes smaller than $\langle \alpha_g \rangle = 0.5$ the parameter $k_c$ in the Eq. (1) becomes fluctuating. Then, in order to calculate the power spectrum in this case one should make use of an ensemble averaging \citep{b}
$$
E(k) \propto \int_0^{\infty} P(k_c) \exp -(k/k_c)dk_c \eqno{(7)}
$$    
with probability distribution $P(k_c)$. \\

  Each of the above considered adiabatic invariants $I_n$ has a corresponding attractor in the phase space. The basins of attraction of these attractors are significantly different: the larger $n$ - the thinner the corresponding basin of attraction (the phenomenon of intermittency). Therefore, the flow dynamics is dominated by the first available invariant $I_n$ having the smallest order $n$. Let us start from  $I_3$ for simplicity (for the case with zero global/net helicity and $I_3 = 0$ see Appendix).                  
  
    The dimensional considerations can be used in order to estimate characteristic velocity $u_c$ for the fluctuating $k_c$ 
 $$
 u_c \propto |I_3|^{1/6} k_c^{1/2}    \eqno{(8)}
 $$
  Assuming a Gaussian (with zero mean) distribution for the characteristic velocity $u_c$  \cite{my}, we obtain from the Eq. (8)
$$
P(k_c) \propto k_c^{-1/2} \exp-(k_c/4k_{\beta})  \eqno{(9)}
$$
where the parameter $k_{\beta}$ is a constant.

\begin{figure} \vspace{-1.5cm}\centering
\epsfig{width=.47\textwidth,file=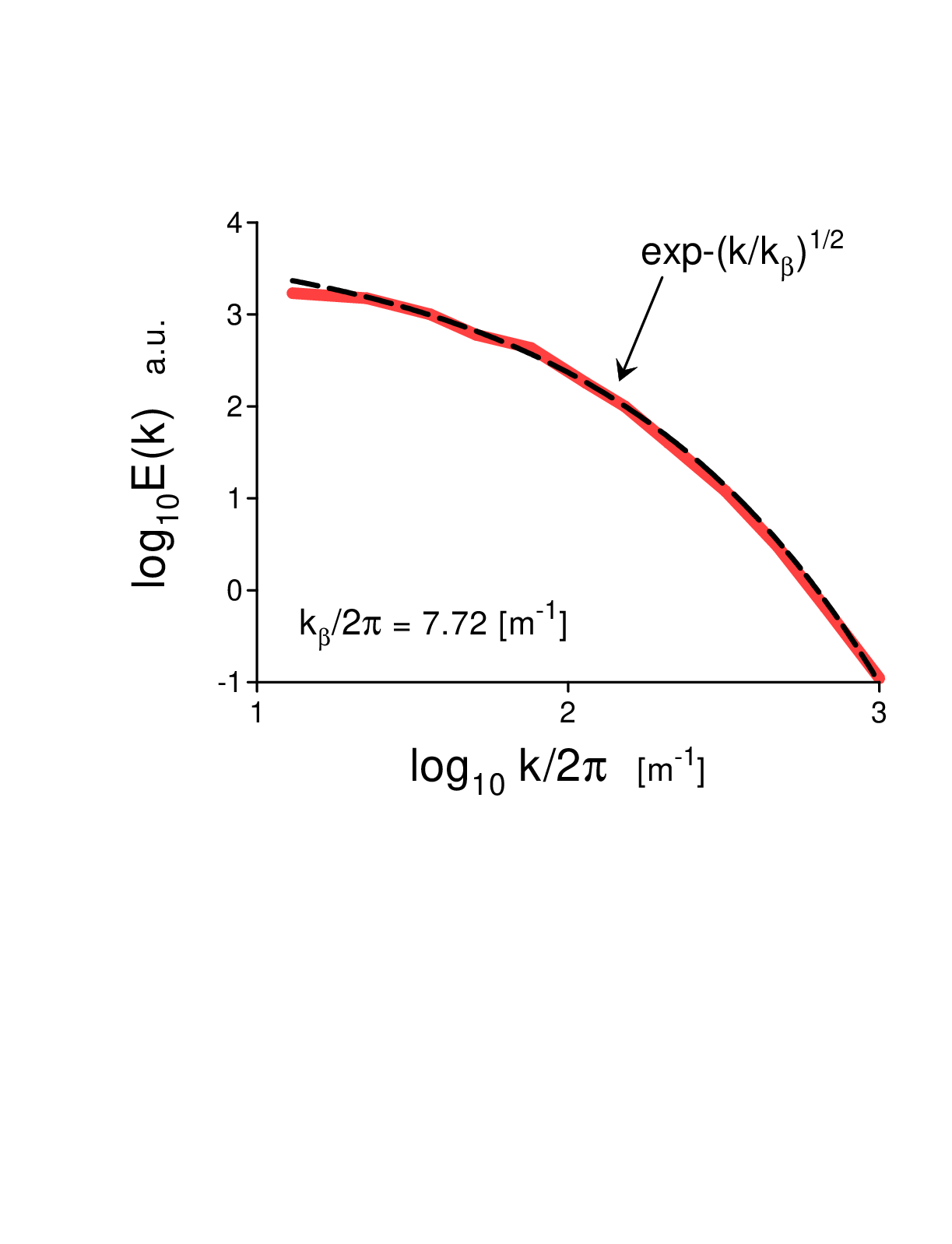} \vspace{-3.9cm}
\caption{Power spectrum of the vertical velocity fluctuations in the liquid phase obtained in the Eulerian-Lagrangian DNS for small value of the  gas volume fraction $\alpha = 0.005$.} 
\end{figure}

    Substituting the Eq. (9) into the Eq. (7) one obtains
$$
E(k) \propto \exp-(k/k_{\beta})^{1/2}  \eqno{(10)}
$$

 \section{Direct numerical simulations and experiments - I}
 
   Let us start from the simplest approach Eqs. (2-3) with the Eulerian-Lagrangian model where the gas phase is composed of spherical non-deformable bubbles which are defined by their centroids position and velocity vectors in initially quiescent water \cite{lai}. Figure 2 shows the power spectrum of the vertical velocity fluctuations of the liquid phase obtained in a direct numerical simulation using this model for a vertical channel with a rising homogeneous swarm of bubbles for a small value of the gas volume fraction $\alpha = 0.005$. The spectral data were taken from Fig. 3a of the Ref. \cite{lai}. The dashed curve indicates the stretched exponential spectrum Eq. (10).\\
   
   This model does not take into account the effects of the bubbles' shape deformation, collisions between bubbles, and the possibility of their coalescence. These effects were taken into account in the `volume-of-fluid' direct numerical simulations, which is a kind of the two-fluid model \cite{lai}. \\
   
   Figure 3 shows the power spectrum of the vertical velocity fluctuations in the liquid phase obtained in a direct numerical simulation using this model for a vertical channel with a rising homogeneous swarm of bubbles for the same small value of the gas volume fraction $\alpha = 0.005$. The spectral data were taken from Fig. 3b of the Ref. \cite{lai}. The dashed curve indicates the stretched exponential spectrum Eq. (10).\\ 
   
	   Results of another two-fluid direct numerical simulation with buoyancy-driven turbulence were reported in the Ref. \cite{ppf}. The power spectrum of the {\it horizontal} velocity fluctuations of the liquid phase obtained in this DNS  at high gas volume fraction $\langle \alpha_g \rangle = 0.5$ was already shown in Fig. 1. Figure 4 shows the power spectrum of the {\it vertical} velocity fluctuations in the liquid phase obtained in this DNS  at the same value of gas volume fraction. The spectral data were taken from Fig. 12b of the Ref. \cite{lai}. The dashed curve indicates the stretched exponential spectrum Eq. (10). One can see that at this high gas volume fraction the horizontal velocity fluctuations are already dominated by the {\it deterministic } chaos (with the exponential spectrum Eq. (1)), while the vertical velocity fluctuations are dominated by the {\it distributed} chaos (with the stretched exponential spectrum Eq. (10)). \\ 
\begin{figure} \vspace{-1.2cm}\centering
\epsfig{width=.45\textwidth,file=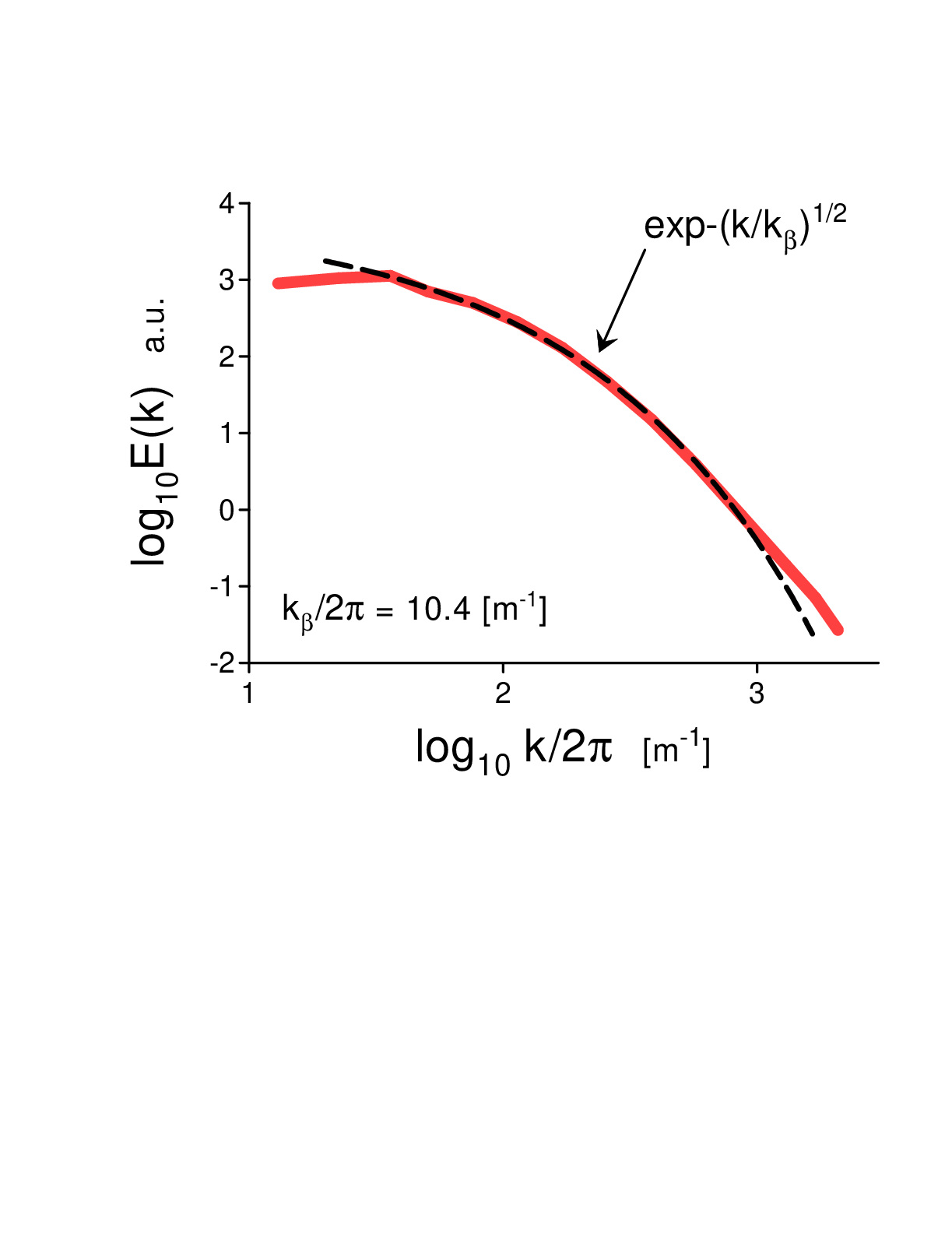} \vspace{-3.9cm}
\caption{Power spectrum of the vertical velocity fluctuations in the liquid phase obtained in the `volume-of-fluid' DNS for small value of the  gas volume fraction $\alpha = 0.005$.} 
\end{figure}
\begin{figure} \vspace{-0.45cm}\centering
\epsfig{width=.45\textwidth,file=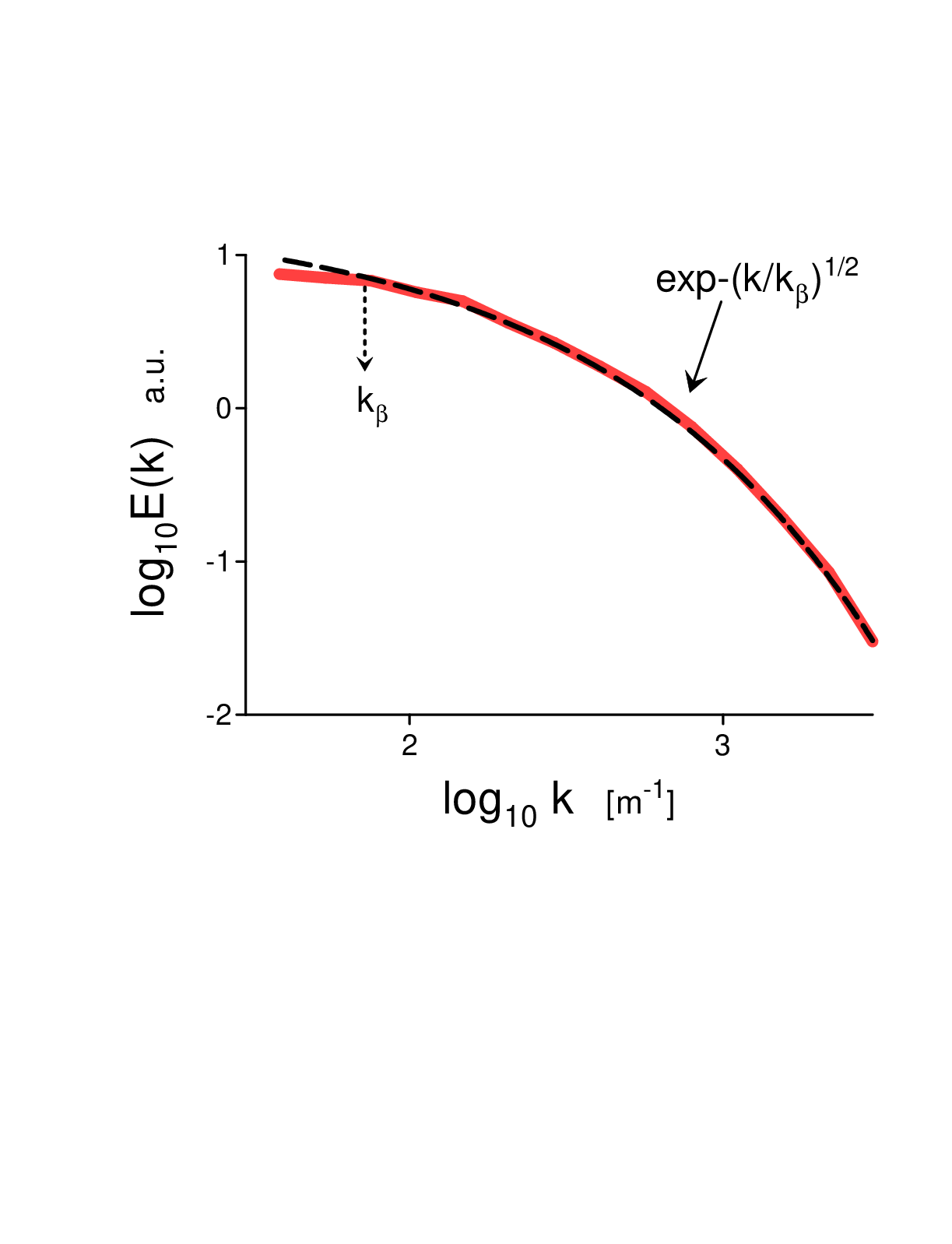} \vspace{-3.6cm}
\caption{Power spectrum of the vertical velocity fluctuations in the liquid phase obtained in the two-fluid DNS for high value of the  gas volume fraction $\langle \alpha_g \rangle = 0.5$.} 
\end{figure}
\begin{figure} \vspace{-0.45cm}\centering
\epsfig{width=.45\textwidth,file=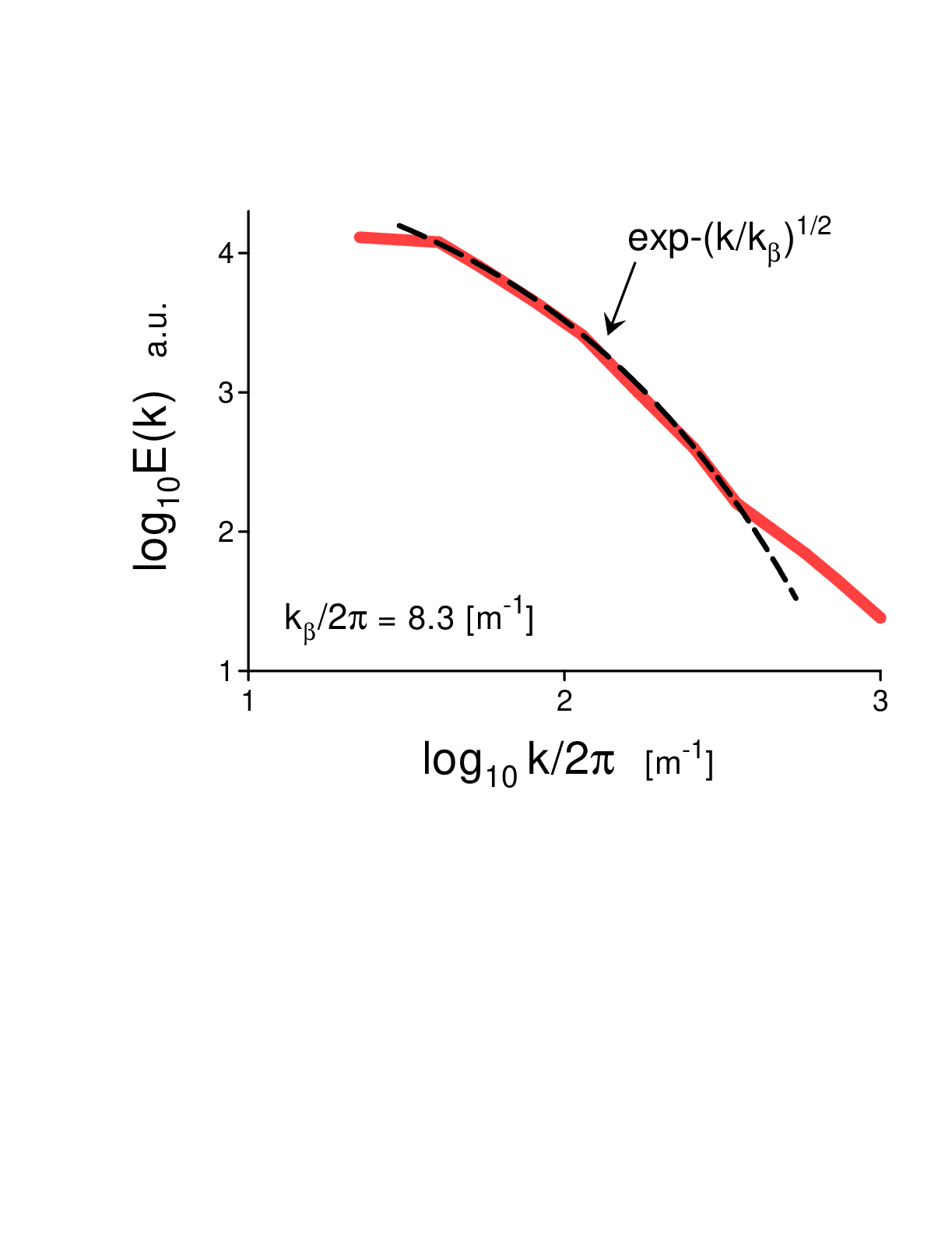} \vspace{-3.9cm}
\caption{ Power spectrum of the horizontal velocity fluctuations in the liquid phase obtained in the experiment \cite{rrl} at moderate gas volume fraction $\alpha = 0.07$.} 
\end{figure}
   
   Let us now turn to the experimental data. Figure 5 shows the power spectrum of the {\it horizontal} velocity fluctuations in the liquid phase obtained in a homogeneous swarm of bubbles (2.5-mm-diameter) rising in water at a moderate gas volume fraction $\alpha =0.07$. The measurements were made 
using laser Doppler anemometry and optical probes \cite{rrl}. Figure 6 shows the power spectrum of the {\it vertical} velocity fluctuations in the liquid phase obtained under the same conditions. The spectral data for Figs. 5 and 6 were taken from Figs. 5b and 5a of the Ref. \cite{ris} correspondingly. The dashed curves indicate the stretched exponential spectrum Eq. (10). One can see that both horizontal and vertical spectra correspond to the distributed chaos type Eq. (10) with the same value of $k_{\beta}$.\\

\section{Variability of the helical distributed chaos} 

\begin{figure} \vspace{-1.5cm}\centering
\epsfig{width=.45\textwidth,file=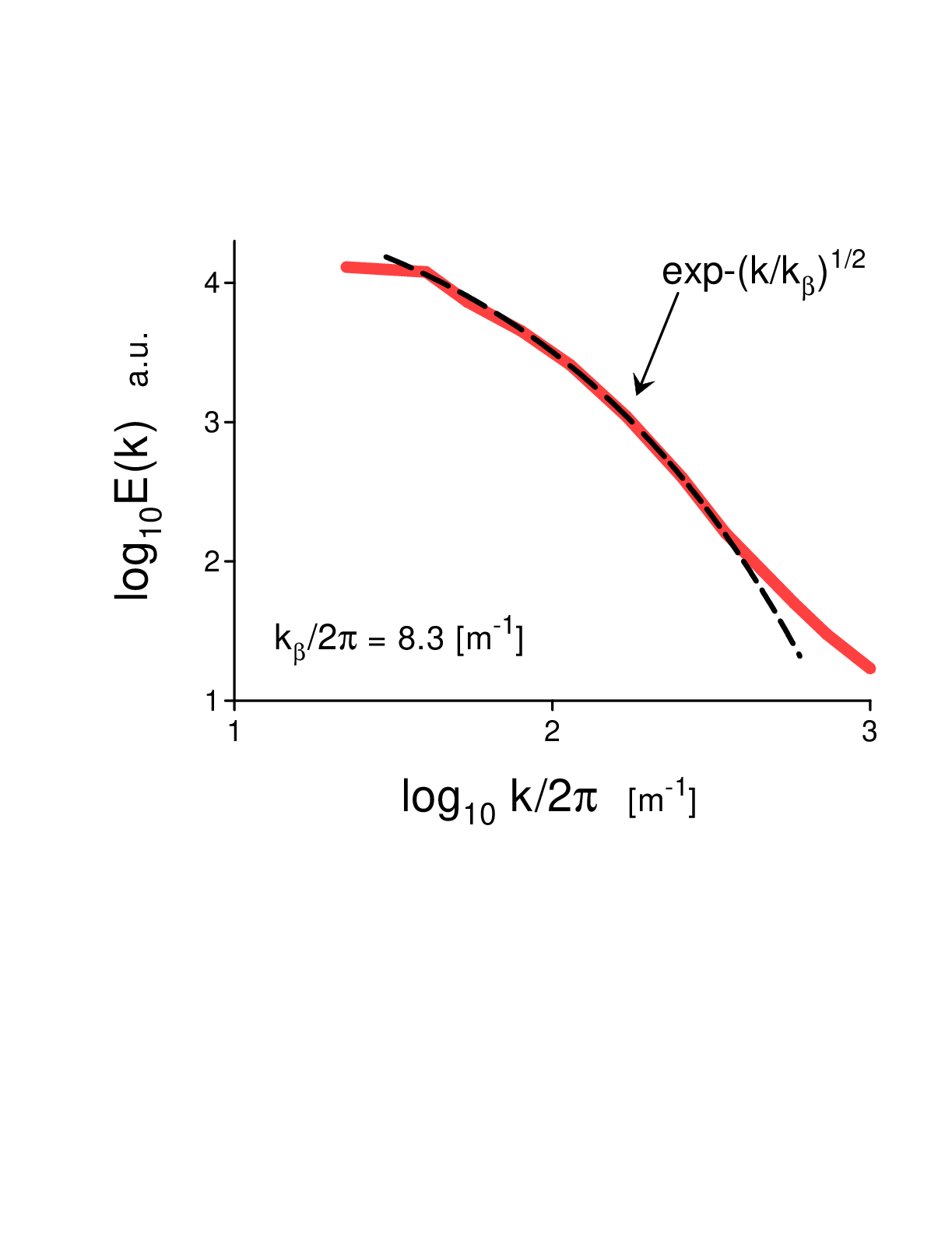} \vspace{-3.5cm}
\caption{The same as in the Fig. 5 but for vertical velocity fluctuations} 
\end{figure}

  Let us consider the general case with the first available (with a minimal $n$) adiabatic invariant $I_n$. The particular estimate Eq. (8) can be replaced by the general estimate obtained from the dimensional considerations
$$
 u_c \propto |I_n|^{1/2n}~ k_c^{\alpha_n}    \eqno{(11)}
 $$    
where 
$$
\alpha_n = 1-\frac{3}{2n}  \eqno{(12)}
$$
Since the straightforward analytic calculations of the ensemble-averaged spectrum cannot be performed in the general case one can use an asymptotic method. One can generalize the stretched exponential spectrum Eq. (10)
 $$
E(k) \propto \int_0^{\infty} P(k_c) \exp -(k/k_c)dk_c \propto \exp-(k/k_{\beta})^{\beta_n} \eqno{(13)}
$$  
where $\beta_n$ and $k_{\beta}$ are some constants. The distribution $P(k_c)$ can be estimated from the Eq. (13) at the asymptotic of large $k_c$
\cite{jon} 
$$
P(k_c) \propto k_c^{-1 + \beta/[2(1-\beta_n)]}~\exp(-\gamma k_c^{\beta_n/(1-\beta_n)}) \eqno{(14)}
$$     
with the $\gamma$ as a constant parameter.\\

 If again the $u_c$ has Gaussian distribution a relationship between the exponents $\beta_n$ and $\alpha_n$ can be readily obtained from the Eqs. (11) and (14)
$$
\beta_n = \frac{2\alpha_n}{1+2\alpha_n}  \eqno{(15)}
$$
 
  Substituting $\alpha_n $  Eq. (12) into the Eq. (15) one obtains
 $$
 \beta_n = \frac{2n-3}{3n-3}   \eqno{(16)}  
 $$
 For $n \gg 1$ the Eqs. (13) and (16) provide
$$
E(k) \propto \exp-(k/k_{\beta})^{2/3}  \eqno{(17)}
$$ 
and for $n=2$
$$
E(k) \propto \exp-(k/k_{\beta})^{1/3}  \eqno{(18)}
$$

\section{Direct numerical simulations and experiments - II}

\begin{figure} \vspace{-1.5cm}\centering
\epsfig{width=.45\textwidth,file=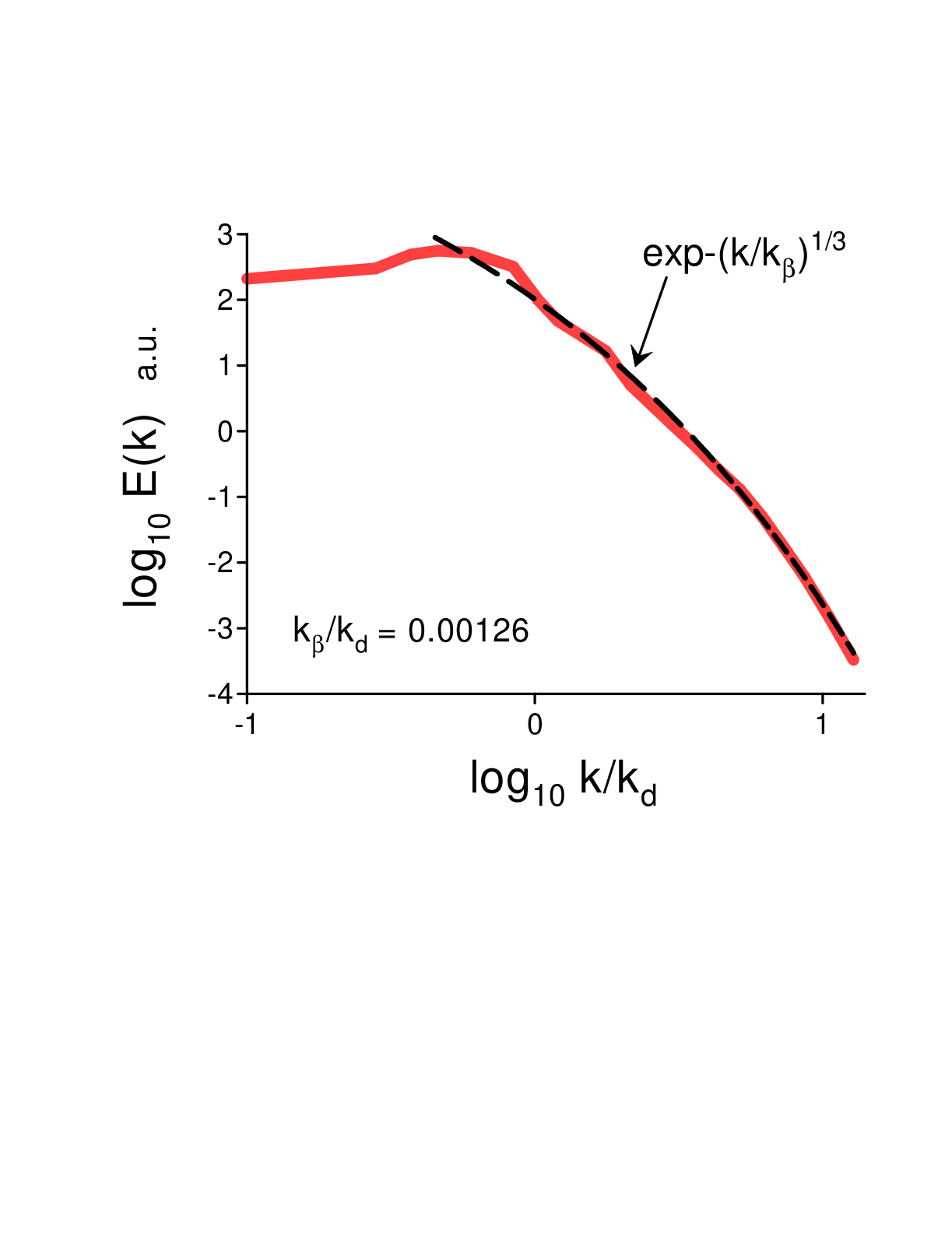} \vspace{-3.5cm}
\caption{Kinetic energy spectrum at $Re=104$ and $At=0.004$.} 
\end{figure}
\begin{figure} \vspace{-0.45cm}\centering
\epsfig{width=.45\textwidth,file=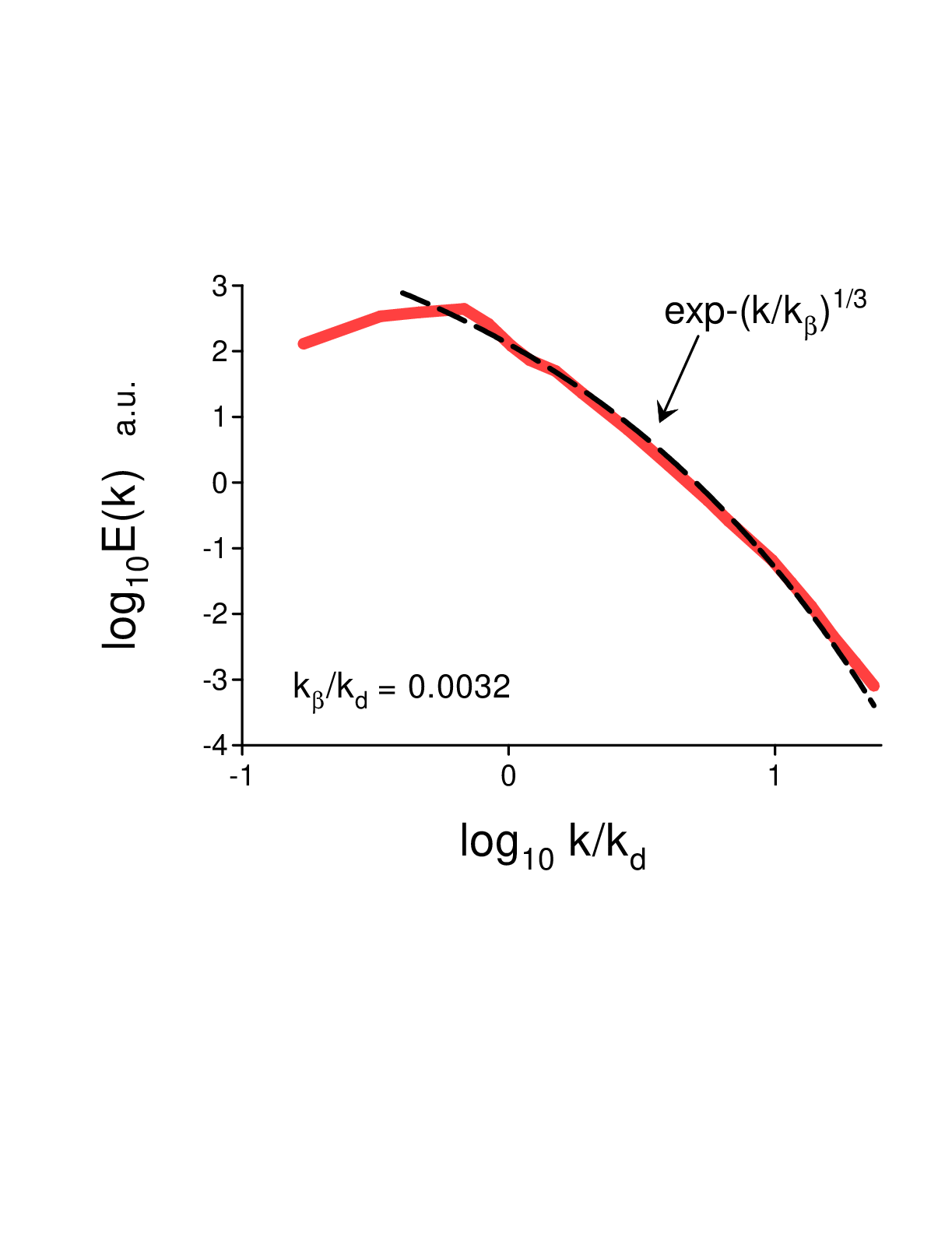} \vspace{-3.1cm}
\caption{As in the Fig. 7 but for $Re=296$.} 
\end{figure}
\begin{figure} \vspace{-1.5cm}\centering
\epsfig{width=.45\textwidth,file=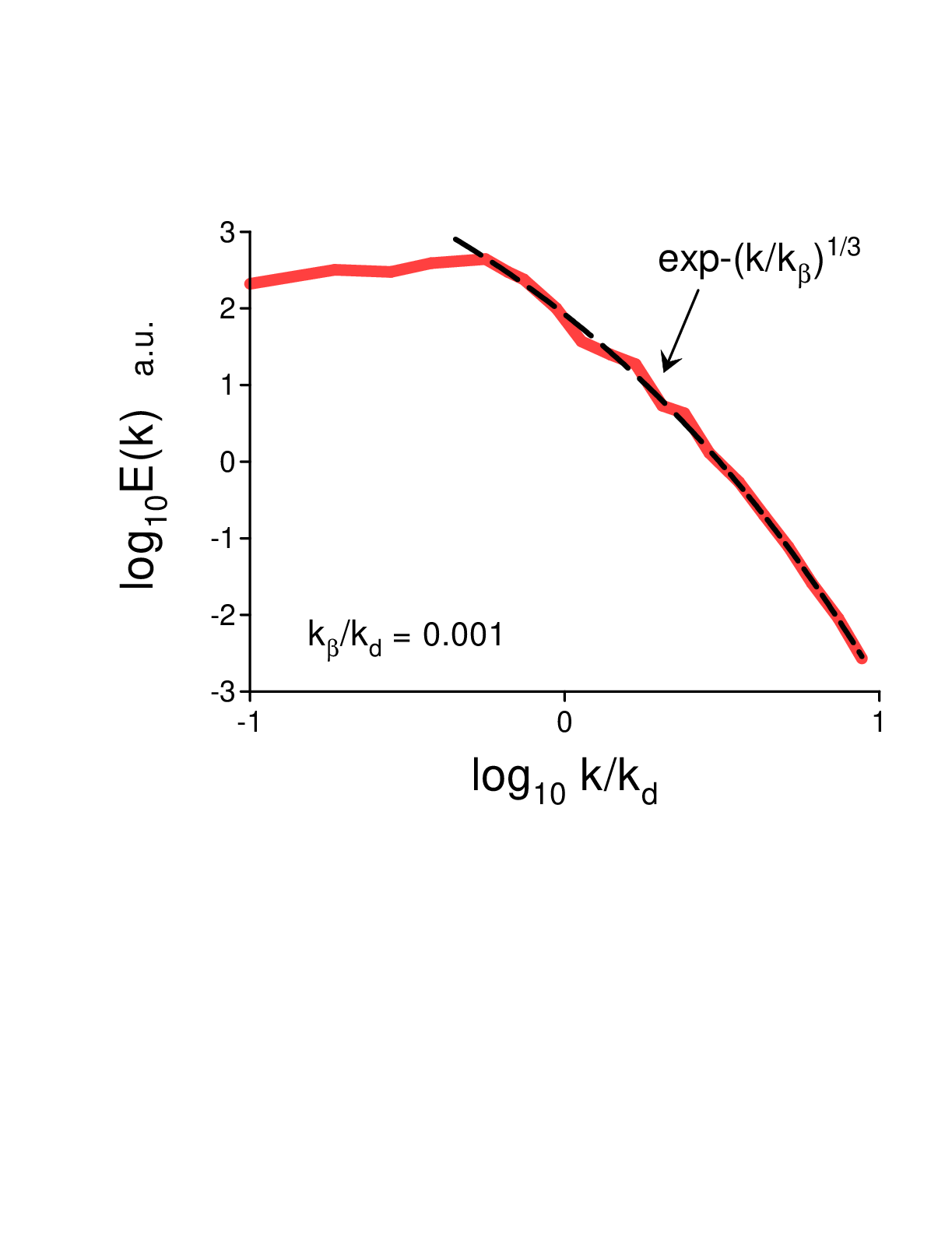} \vspace{-3.5cm}
\caption{Kinetic energy spectrum at $Re=113$ and $At=0.90$.} 
\end{figure}
\begin{figure} \vspace{-0.25cm}\centering
\epsfig{width=.45\textwidth,file=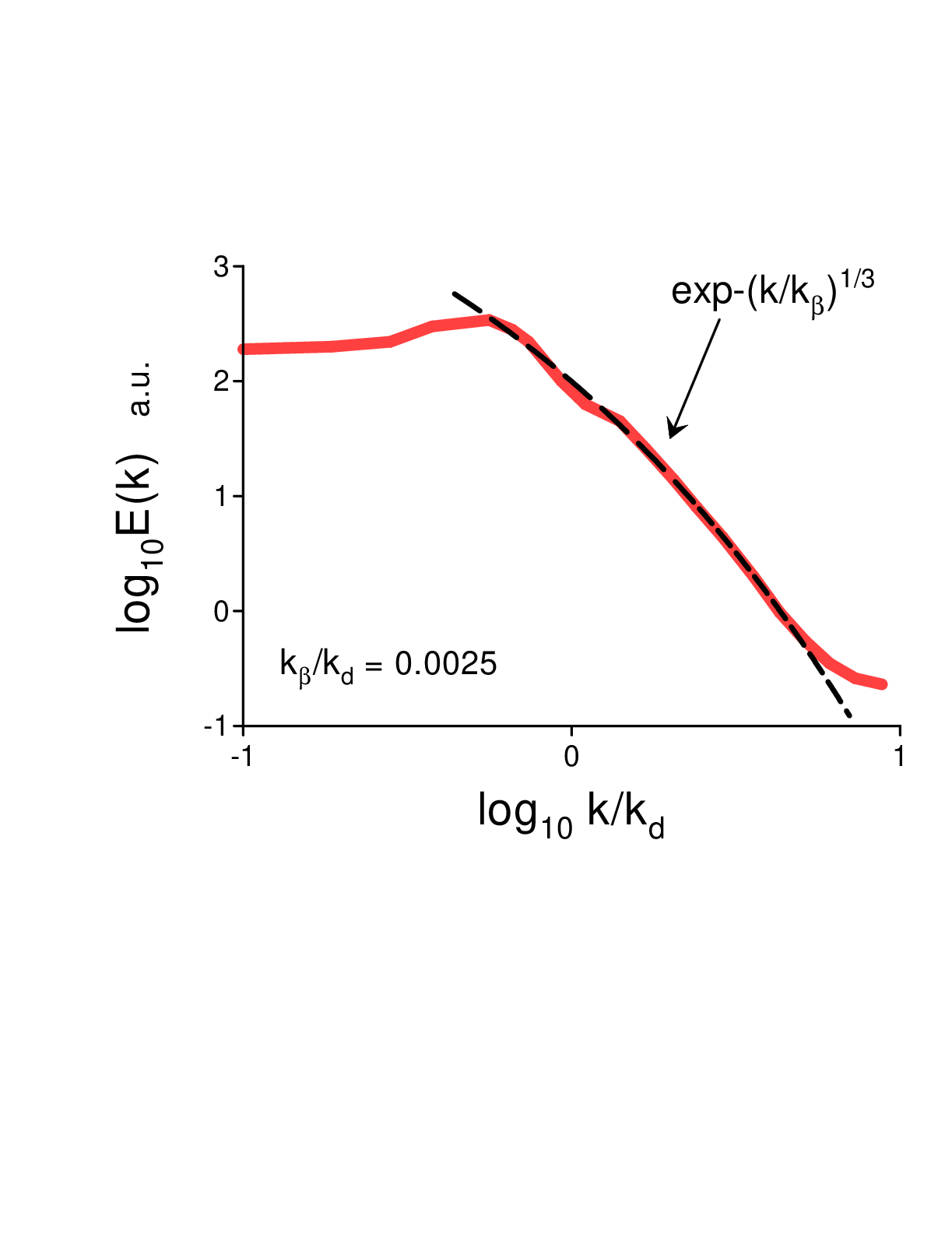} \vspace{-3.2cm}
\caption{As in the Fig. 9 but for $Re=358$.} 
\end{figure}
   In a recent paper Ref. \cite{prp} results of direct numerical simulations of the pseudo-turbulence in the buoyancy-driven bubbly flow were presented for different values of the Atwood and Reynolds numbers. In these numerical simulations the Navier-Stokes equations 
 $$
 \rho [\frac{\partial {\bf u}}{\partial t} + ({\bf u} \cdot \nabla) {\bf u} ] =  -\nabla p +  \nabla \cdot [2\mu \mathcal{S}]  +{\bf F}  \eqno{(19)}
$$
$$
\nabla \cdot {\bf u} =  0 \eqno{(20)}
$$
were studied with the density $\rho = \rho_f c+\rho_b (1-c)$, the viscosity $\mu = \mu_f c+ \mu_b (1-c)$ and the deformation rate tensor $\mathcal{S}$. The body force term 
$$
{\bf F} = (\langle \rho \rangle-\rho) g {\bf e_z} + \sigma \kappa {\bf n}  \eqno{(21)}
$$
 where $g$ is the gravity acceleration, ${\bf e_z}$ is a unit vector in the vertical direction, ${\bf n}$ is the unit normal vector to the bubble interface, $\kappa$ is the curvature of the bubble surface, $\sigma$ is surface tension coefficient.  The first term in the right-hand side of the Eq. (21) represents the buoyancy force and the second term represents the surface tension. The DNS were performed in a cubic domain with periodic boundary conditions. 
 
    Figure 7 shows the kinetic energy spectrum for Reynolds number (constructed using the bubble diameter $d$) $Re = [\rho_f (\rho_f-\rho_b)gd^3]^{1/2}/\mu_f = 104$ and Atwood number $At = (\rho_f-\rho_b)/( \rho_f+\rho_b) = 0.004$. The spectral data were taken from Fig. 5a of the Ref. \cite{prp} ($k_d$ is the wavenumber corresponding to the bubble diameter). The dashed curve is drawn to indicate correspondence to the spectrum Eq. (18). Figure 8 shows analogous spectrum but obtained for $Re = 296$.
    
\begin{figure} \vspace{-1.55cm}\centering
\epsfig{width=.45\textwidth,file=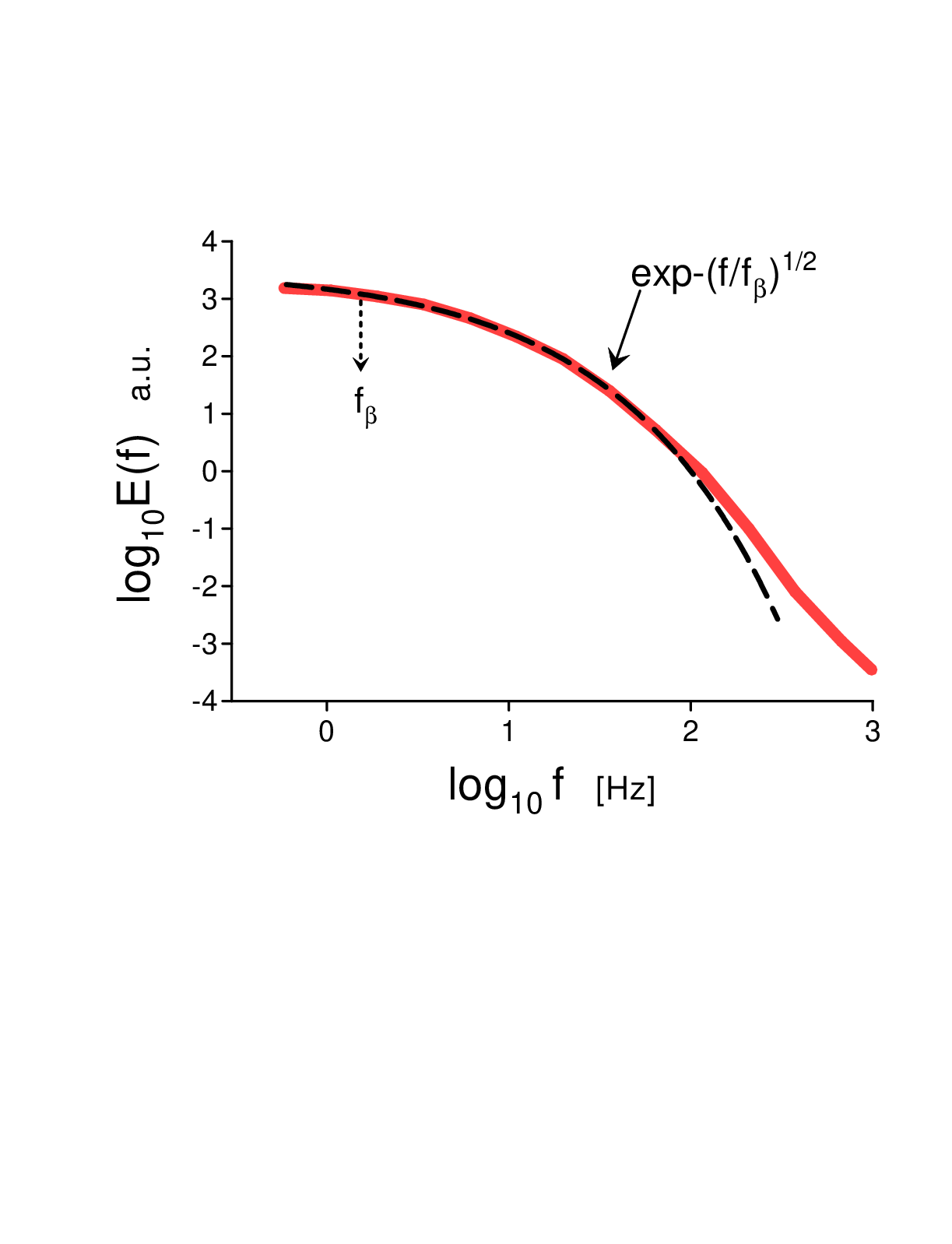} \vspace{-3.5cm}
\caption{Power spectrum of the vertical velocity fluctuations for the set 1 (`large' bubbles) at $b = \infty$ and $\alpha \simeq 0.01$. } 
\end{figure}
\begin{figure} \vspace{-0.45cm}\centering
\epsfig{width=.45\textwidth,file=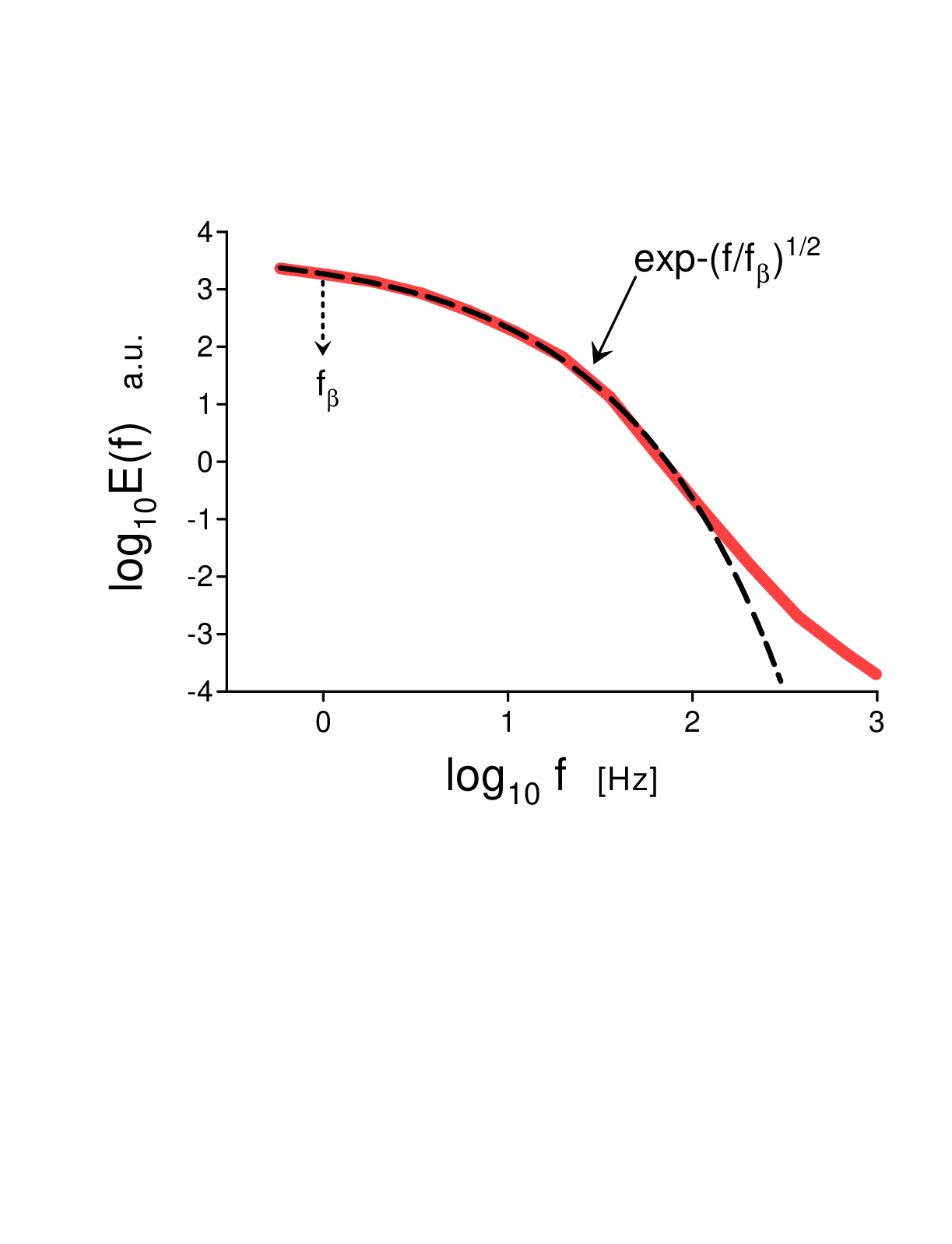} \vspace{-3.5cm}
\caption{As in the Fig.11 but for $b \simeq 1$.} 
\end{figure}

     Figure 9 shows the kinetic energy spectrum for Reynolds number $Re =113$ and Atwood number $At = 0.90$. The spectral data were taken from Fig. 6a of the Ref. \cite{prp}. The dashed curve is drawn to indicate correspondence to the spectrum Eq. (18). Figure 10 shows analogous spectrum but obtained for $Re = 358$.\\
     
     In Ref. \cite{prak} results of an experiment performed for the bubbly flows in a quiescent liquid (a vertical water tunnel) and in the presumably isotropic homogeneous turbulence generated by an active grid in this water tunnel were reported. The rising bubbles were produced by capillary islands at the bottom of the tunnel. Two types of capillary needles were used in order to obtain two sets of bubble sizes: set 1 with the inner diameter of the needles equal to 500$\mu$m and set 2 with the inner diameter of the needles equal to 120$\mu$m. The bubbles in set 1 had a diameter $ 3<d<5$mm whereas in set 2 the diameter was $2<d<4$mm. The mean velocity of the flow was in the upward direction. \\ 
     
\begin{figure} \vspace{-1.5cm}\centering
\epsfig{width=.45\textwidth,file=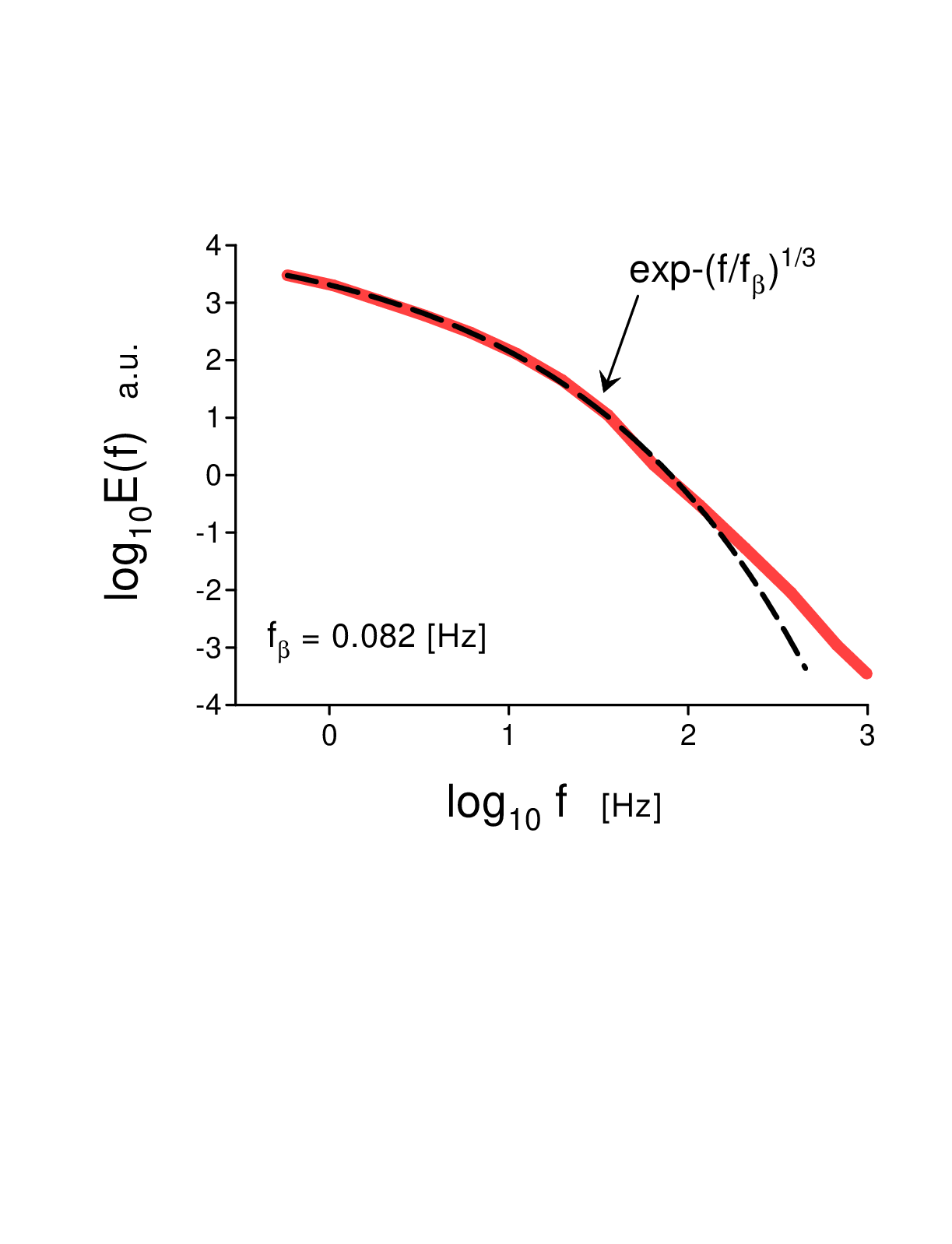} \vspace{-3.5cm}
\caption{Power spectrum of the vertical velocity fluctuations for the set 2 (`small' bubbles) at $b = \infty$ and $\alpha \simeq 0.01$. } 
\end{figure}
\begin{figure} \vspace{-0.5cm}\centering
\epsfig{width=.45\textwidth,file=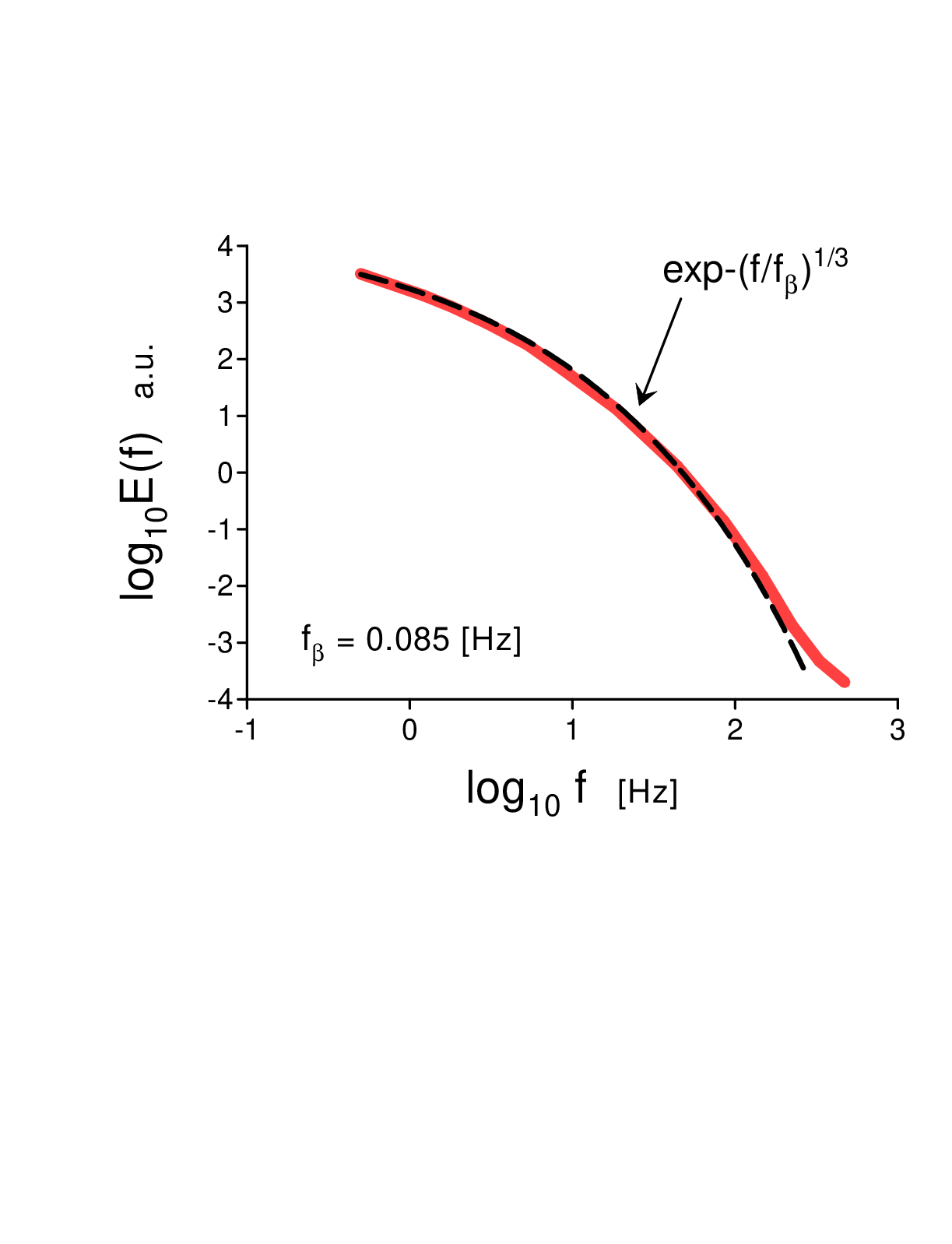} \vspace{-3.5cm}
\caption{As in the Fig.13 but at $b \simeq 1$.} 
\end{figure}
\begin{figure} \vspace{-1.35cm}\centering
\epsfig{width=.45\textwidth,file=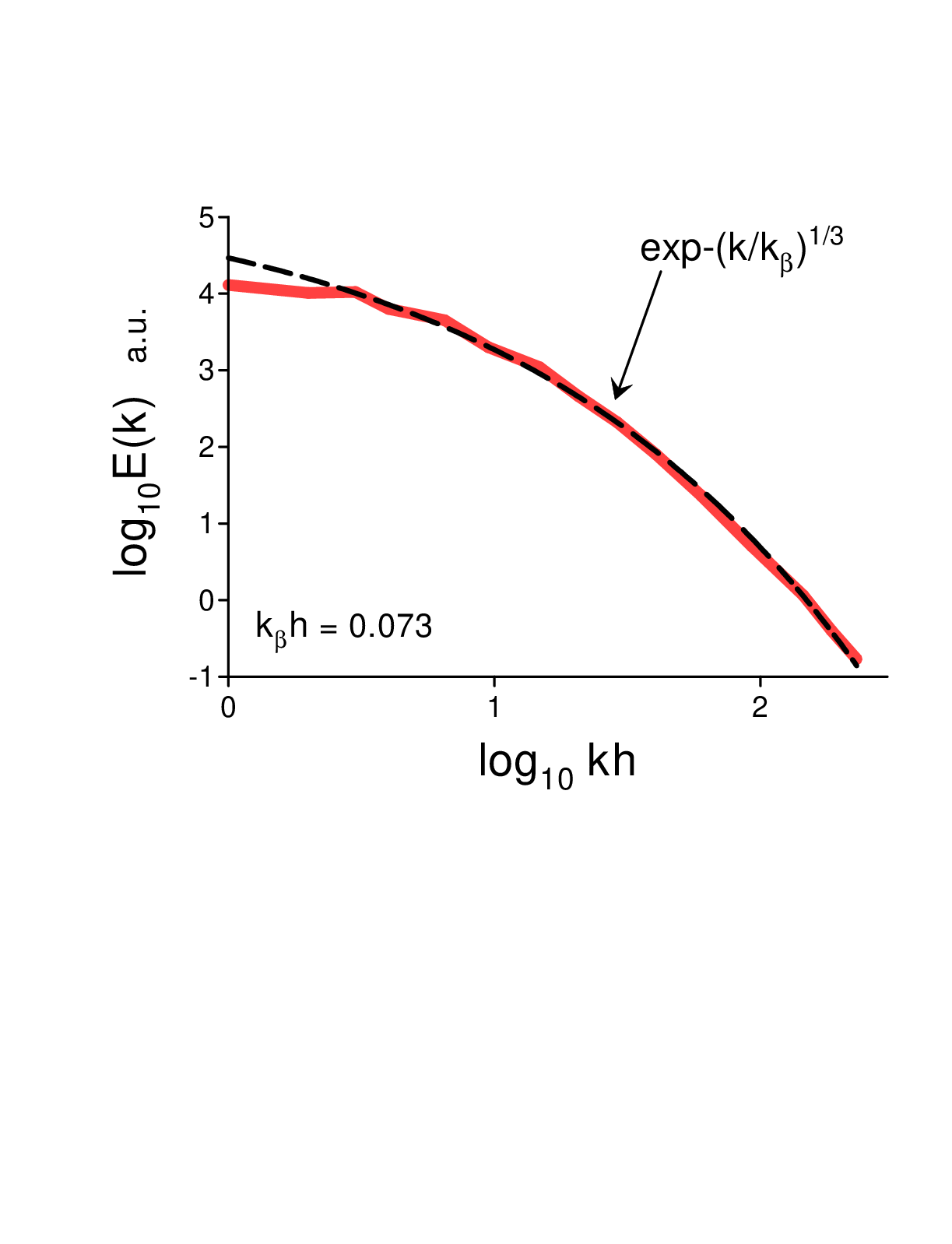} \vspace{-3.65cm}
\caption{Power spectrum of the streamwise liquid velocity fluctuations for the near-wall bubbly flow. } 
\end{figure}
    
     In order to characterize activity of the background flow the authors of the Ref. \cite{prak} used the so-called bubblance parameter \cite{lb},\cite{rll}:
 $$
 b = \frac{1}{2}\frac{\alpha U_r^2}{u_0^2}      \eqno{(22)}
 $$
 $U_r$  is the velocity of the bubble rising in still water, $u_0^2$ is typical turbulent fluctuations of the fluid velocity when the bubbles are absent. For single-phase flows $b=0$ and $b \simeq \infty$ for the case when the velocity fluctuations are mainly caused by bubbles.\\

     Figures 11 and 12 show the power spectra of the vertical velocity fluctuations measured in the experiment for the set 1 (`large' bubbles) but with different values of $b$ (for Fig. 11 - $b = \infty$ and $\alpha \simeq 0.01$, whereas for Fig. 12 - $b \simeq 1$).  The spectral data were taken from Figs. 6a and 6c of the Ref. \cite{prak} respectively. The Taylor hypothesis relating the wavenumber ($k$) and frequency ($f$) spectra using relationship  $f =Uk/2\pi$ (where $U$ is the mean velocity of the flow) can directly relate the power spectra  $E(f) = E(k)~2\pi/U$ (see, for instance, \cite{kv} and references therein).  Therefore, the dashed curves are drawn to indicate correspondence to the spectrum Eq. (10) dominated by the third moment of the helicity distribution.  \\
     
       Figures 13 and 14 show the power spectra of the vertical velocity fluctuations measured in the experiment for the set 2 (`small' bubbles) but with different values of $b$ (for Fig. 13 - $b = \infty$ and $\alpha \simeq 0.01$, whereas for Fig. 14 - $b \simeq 1$).  The spectral data were taken from Figs. 6b and 6d of the Ref. \cite{prak} respectively. The dashed curves are drawn to indicate correspondence to the spectrum Eq. (18) dominated by the second moment of the helicity distribution.\\ 
   
       Another example of interaction between bubbles and a background near-wall turbulent flow was presented in a recent paper Ref. \cite{mrl}. This Large-Eddy simulation was performed in a convective air-water channel flow in order to simulate bubbly near-wall turbulence at the fuel rods in the pressurized water reactors' hot channels. Such reactors are widely used at nuclear power plants. In the streamwise and spanwise directions of the channel the periodic boundary conditions were used in this simulation. It is well known that the small vapor bubbles appearing at  the fuel rods and then departing into the near-wall turbulence significantly alter the flow and, consequently, the heat and mass transfer. When the near-wall turbulence reached a state which is usually considered a fully developed one (at $Re_{\tau} =400$) about 600 small air bubbles were introduced in this simulation at the wall's surface. Detachment of the bubbles, their deformation and transport were allowed in close proximity to the corresponding wall. About 50\% of the bubbles were detached from the wall at the end of the simulation.\\
       
       Figure 17 shows a power spectrum of the streamwise liquid velocity fluctuations at the dimensionless height $y^+ =20$ above the wall surface. The spectral data were taken from Fig13a of the Ref. \cite{mrl} for the bubbly flow ($h$ is the half channel height). The dashed curve is drawn to indicate correspondence to the spectrum Eq. (18) dominated by the second moment of the helicity distribution.
       
\section{Appendix}

  Not only the spectrum Eq. (18) but also the spectrum Eq. (10) can be valid in the case when the global/net helicity and $I_3$ are equal to zero due to global parity (mirror) symmetry.  It is shown in the paper Ref. \cite{ker} (using numerical simulations) that a spontaneous local breaking of the parity symmetry, due to inviscid vortex interactions, can result in the appearance of a pair of adjacent local regions with strong helicity. Because of opposite signs of the helicity in these regions the global helicity is remained equal to zero due to the global symmetry. This mechanism is inherent to the turbulent flows and can be modified in the complex turbulent flows with bubbles, but the result should be the same. \\
  
   Let us denote the helicity of the cells with positive helicity $H_j^{+}$ and with negative helicity $H_j^{-}$, and 
$$
I_3^{\pm} = \lim_{V \rightarrow  \infty} \frac{1}{V} \sum_j [H_{j}^{\pm}]^3  \eqno{(A1)}
$$ 
where the summation is made for the positive (or negative) helicity cells only.  

   Due to the global symmetry $I_3 = I_3^{+} + I_3^{-} =0$ and consequently $I_3^{+} = - I_3^{-}$. Applying the same consideration as in the Section II one can conclude that $|I_3^{\pm}|$ can be still a finite (non-zero) ideal quasi-invariant and the estimate Eq. (8) can be replaced by the estimate
 $$
 u_c \propto |I_3^{\pm}|^{1/6} k_c^{1/2},    \eqno{(A2)}
 $$
and the kinetic energy spectrum has the same form Eq. (10) in the case of the global symmetry as well.

\end{document}